\documentclass[12pt]{iopart}

\usepackage{graphicx}% Include figure files
\usepackage{dcolumn}% Align table columns on decimal point
\usepackage{bm}% bold math
\usepackage{epsfig}
\usepackage{color}
\usepackage{bbm}
\usepackage{longtable}
\usepackage{graphicx}% Include figure files
\usepackage{dcolumn}% Align table columns on decimal point
\usepackage{bm}% bold math
\usepackage{iopams}
\newcommand{\be}{\begin{equation}}
 \newcommand{\ee}{\end{equation}}
\newcommand{\bear}{\be\begin{array}}
\newcommand{\bea}{\begin{eqnarray}}
\newcommand{\eea}{\end{eqnarray}}

\newcommand{\bR}{{\bf R}}

\def\ketm#1{  \left\vert  #1   \right\rangle   }

\def\mem#1#2#3{  \left\langle #1 \left\vert  #2 \right\vert #3 \right\rangle   }

\def\asc#1{{\color{black}#1}}                 % corrected text
\begin{document}

%\preprint{}
%
%
%-----------------------------------------Title---------------------------------------
%
%
\title{Coulomb excitation of hydrogen atoms by vortex ion beams}
%
%
%-----------------------------------------Author--------------------------------------
%
%
\author{A V Maiorova$^{1,2}$, D Karlovets$^{3}$, S Fritzsche$^{1,2,4}$, A Surzhykov$^{5,6,7}$
Th St\"ohlker$^{1,2,8}$}

\address{$^1$ GSI Helmholtzzentrum für Schwerionenforschung, Planckstra{\ss}e 1, D-64291 Darmstadt, Germany}
\address{$^2$ Helmholtz Institute Jena, Fr\"obelstieg 3, D-07743 Jena, Germany}
\address{$^3$ Qingdao Innovation and Development Center, Harbin Engineering University, Qingdao, Shandong, 266000 China}
\address{$^4$Theoretisch-Physikalisches Institut, Friedrich-Schiller-Universität Jena, D-07743 Jena, Germany}
\address{$^5$ Physikalisch--Technische Bundesanstalt, Bundesallee 100, D--38116 Braunschweig, Germany}
\address{$^6$ Institut f\"ur Mathematische Physik, Technische Universit\"at Braunschweig, Mendelssohnstra{\ss}e 3, D--38106 Braunschweig, Germany}
\address{$^7$ Laboratory for Emerging Nanometrology Braunschweig, Langer Kamp 6a/b, D-38106 Braunschweig, Germany}
\address{$^8$Institut für Optik und Quantenelektronik, Friedrich-Schiller-Universit\"at Jena, D-07743 Jena, Germany}
\ead{A.Maiorova@gsi.de}

\date{\today \\[0.3cm]}

%
%
%-----------------------------------------Abstract--------------------------------------
%
%

\begin{abstract}
Coulomb excitation of hydrogen atoms by vortex protons is theoretically investigated within the framework of the non--relativistic first--Born approximation and the density matrix approach. Special attention is paid to the magnetic sublevel population of excited atoms and, consequently, to the angular distribution of the fluorescence radiation. We argue that both these properties are sensitive to the projection of the orbital angular momentum (OAM), carried by the projectile ions. In order to illustrate the OAM--effect, detailed calculations have been performed for the $1s \to 2p$ excitation and the subsequent $2p \to 1s$ radiative decay of a hydrogen target, interacting with incident Laguerre--Gaussian vortex protons. The calculation results suggest that Coulomb excitation can be employed for the diagnostics of vortex ion beam at accelerator and storage ring facilities.
\end{abstract}

\maketitle

%
% ------------------------------------------------------- Introduction -----------------------------------------------------------------------
%
\section{Introduction}

Investigations of vortex, or twisted, states of light and matter have a rich history of more than thirty years. These states, which possess a helical phase front and carry a non--zero projection $l_{OAM}$ of the orbital angular momentum (OAM) onto their propagation direction, were first predicted and experimentally realized for the \textit{photon} beams in the early nineties of the last century \cite{Allen1992,He1995}. Since then, vortex light has found its applications in many areas of modern physics, ranging from super--resolution microscopy and magnetomery to classical and quantum information processing, operation of atomic clock transitions and optical tweezing \cite{HeFr95,Ber2009,BarWe2008,Padgett2017,Shen2019, Babiker_2019}. Following the success of twisted photon research, Bliokh and co–-workers in 2007 \cite{Bli07} predicted and described the properties of free \textit{electron} states with non--zero OAM projection and phase vorticies. Having been produced just a few years later \cite{UcT10,Verbeeck2010,McA11,MafakAPL17}, the OAM electrons rapidly demonstrated their unique potential for transmission electron microscopy (TEM), probing magnetic properties of materials and for studying light--matter coupling  \cite{Verbeeck2010,Lloyd2012,Verbeeck2013,Harris2015}. The next particles produced in the vortex state were \textit{neutrons} \cite{Cla2015,Larocque2018}. This opens up fundamentally new possibilities for quantum tomography of neutron states, as well as for studies in hadronic and low--energy nuclear physics. Finally, the successful experimental realization of vortex \textit{atoms} and even \textit{molecules} was reported in 2021 \cite{Luski2021}, providing a novel tool for investigations of fundamental interactions, chirality and parity violation. 

A natural extension of the above-mentioned research of vortex states of light and matter is positively--charged \textit{ions} in accelerator and storage ring facilities. The OAM projection of vortex ions will deliver an additional and very intriguing degree of freedom for collision studies in atomic and nuclear physics. However, the realization of these collision experiments requires the development of techniques for production and diagnostics of vortex ions. For the former task, two approaches were proposed recently that are based on the use of (i) forked diffraction gratings, similar to those in Ref.~\cite{Luski2021}, or (ii) magnetized stripping foils \cite{Karl2021}. Both methods  can generate low--intensity Laguerre--Gaussian (LG) ion beams with the OAM projection $l_{OAM}$ and the transverse coherence length \asc{from about 0.1 $\mu$m and larger.}  To verify the non--zero $l_{OAM}$ of the produced vortex beams, their intensity profile can be measured, as was demonstrated for neutral atoms in Ref.~\cite{Luski2021}. 
An alternative and very promising approach for the OAM--diagnostics is based on the analysis of fundamental atomic processes, induced in collisions with vortex ions. In the present work, we discuss the feasibility of this approach for the example of Coulomb excitation and subsequent radiative decay of target atoms. A number of studies of such a two step ``excitation and decay'' process have been performed in the past for conventional (plane--wave) projectiles  \cite{Blu12,LiS89,AnD90,EICHLER2005vii}. These studies have demonstrated that angular distribution and polarization of the decay radiation uniquely reflect the magnetic sublevel population of excited target atoms and, consequently, details of collisions. Here, we will extend theoretical studies of Coulomb excitation towards Laguerre--Gaussian projectiles to demonstrate the sensitivity (of the properties) of this process to the OAM projection.

In our study, we will focus on the $1s \to 2p$ excitation of hydrogen atoms, bombarded by Laguerre--Gaussian protons, $\mathrm{H}(1s) + \mathrm{H}^{+}_{LG}\rightarrow \mathrm{H}(2p) + \mathrm{H}^{+}$. This process can be naturally described in the framework of non--relativistic first--Born approximation, whose basic ideas and formulas are presented in Section~\ref{subsec:Excitation amplitude}. By making use of Born transition amplitudes and the density matrix approach we derive in Section~\ref{subsec:Cross section} the total excitation probability and the alignment parameter, that describe the magnetic sublevel population of the $2p$ hydrogenic state. In Section~\ref{subsec:Angular distribution} both of these parameters are used to predict the angular distribution of the $2p \to 1s$ radiation for both the case of a single H atom and for a \textit{macroscopic} hydrogen target. For the latter case, which is of particular interest for future experiments, it is shown that the angular distribution of fluorescence photons is indeed sensitive to the OAM projection of incident protons. Detailed calculations supporting this conclusion, and hence the practical applicability of Coulomb excitation for the diagnostics of vortex ion beams, are presented in Section~\ref{sec:results}. Finally, the paper is concluded by a short summary in Section~\ref{sec:outlook}.

% ------------------------------------------- Theory ----------------------------------------------- %
%
\section{Theory}
\label{sec:theory}

\subsection{Excitation amplitude}
\label{subsec:Excitation amplitude}

\begin{figure}[t]
   \includegraphics[width=0.8\linewidth]{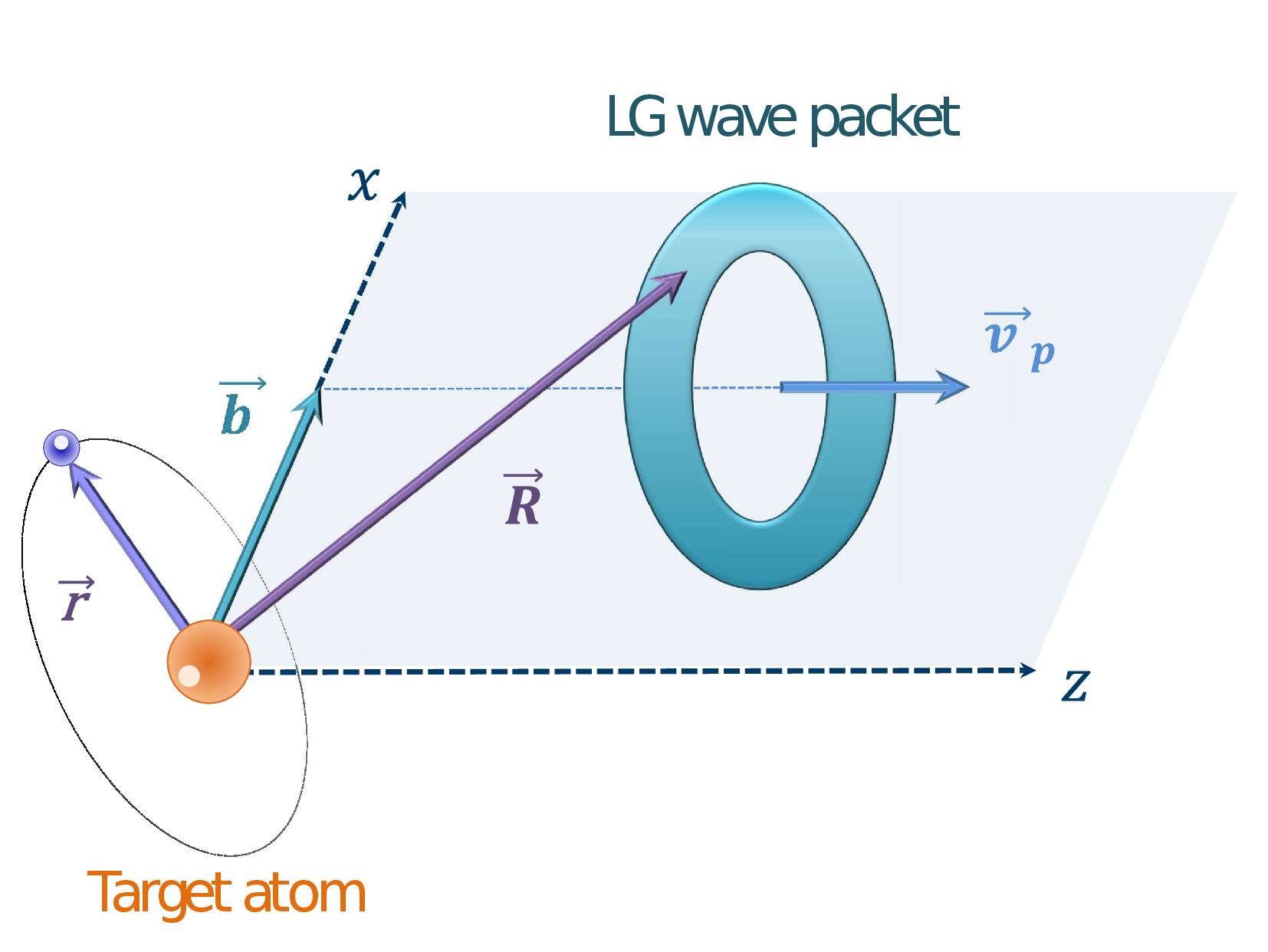}
%	\includegraphics[width=0.99\linewidth]{Fig1.pdf}
	%\vspace*{-2cm}
	\caption{Geometry of Coulomb excitation of the hydrogen-like atom by the vortex ion beam representing the Laguerre-Gaussian wave packet.
	}
    \label{Fig1}
\end{figure}

In the present work we consider the excitation of a hydrogen--like ion from an initial $|n_i\,l_i\,m_i\rangle$ to a final state $|n_f\,l_f\,m_f\rangle$ due to the Coulomb interaction with a projectile ion. We assume that the ion  is prepared in a Laguerre--Gaussian (LG) state \cite{Karl2018, Karl2021}, which is \textit{an exact solution} of the free time-dependent Schr\"odinger equation and whose properties will be discussed below. For theoretical analysis of the excitation process, it is convenient to choose the quantization axis (\emph{z}-axis) of the entire system parallel to the mean velocity $\mathbf{v_p} = (0,0,v_p)$ of the ionic wave packet, see Fig. \ref{Fig1}. Together with the target atom, that we place at the origin, the direction of the mean velocity of the twisted beam defines the reaction plane (xz-plane), wherein the center of the LG wave packet is shifted from the atom by an impact parameter $\mathbf{b} = (b,0,0)$. The position of the atomic electron relative to the target nucleus is determined by the vector $\mathbf{r}$, while the vector $\mathbf{R} = (R_x,R_y,R_z) \equiv (\mathbf{R}_{\bot}, R_z)$ characterizes the charge density of the wave packet as ``seen'' from the target, see Fig.~\ref{Fig1}.

%
% ------------------------------------ Figure 1 -------------------------------------------

%

After discussing the geometry and basic notations, we are ready to proceed with the theoretical analysis of the Coulomb excitation process. Within the non--relativistic framework, the Hamiltonian of a hydrogen atom, interacting with a charged projectile, can be written in the form:
\begin{eqnarray}
   \label{eq:hamiltonian}
\hat{H} = -\frac{\hbar^2}{2m}\nabla^2-\frac{e^2}{r} + V(\mathbf{r},t) \equiv \hat{H}_0 + \hat{V}(\mathbf{r},t).
\end{eqnarray}
Here, the first two terms represent the usual (unperturbed) Hamiltonian of the hydrogen atom, while the perturbation is given by the interaction
\begin{eqnarray}
   \label{eq:intoperator}
\hat{V}(\mathbf{r},t) = -Z_{p}\,e^{2}\int\frac{\rho\,(\mathbf{R},t)\,\mathrm{d} \mathbf{R}}{|\mathbf{r}-\mathbf{R}|},
\end{eqnarray}
between the bound electron and the ionic wave packet with charge $Z_p$ and time--dependent (normalized) charge density $\rho\,(\mathbf{R},t)$. Thus, our approach resembles the well established semi--classical theory of atomic collisions, in which a projectile moves along a classical trajectory.

For the case of a non--relativistic Laguerre--Gaussian wave packet with the orbital angular momentum $l_{OAM}$ and a principal quantum number $n=0$, the density $\rho\,(\mathbf{R},t)$ can be written in the following form \cite{Karl2021,Karl2018,LuiIvanov23}:
\begin{eqnarray}
   \label{eq:chargedensity}
\rho\,(\mathbf{R},t) &\equiv& \left| \Psi^{LG}(\mathbf{R} - \mathbf{b},t)\right|^2 \nonumber \\ 
&& \hspace*{-4cm} = \frac{1}{\sqrt{\pi^3}\,|l_{OAM}|!}\frac{1}{\sigma_z(t)\,\sigma^2_{\bot}(t)} \left(\frac{\left(\mathbf{R}_{\bot}-\mathbf{b}\right)^2}{\sigma^2_{\bot}(t)}\right)^{|l_{OAM}|} \exp\left[-\frac{\left(\mathbf{R}_{\bot}-\mathbf{b}\right)^2}{\sigma^2_{\bot}(t)}-\frac{\left(R_{z}-v_{p}\,t\right)^2}{\sigma^2_{z}(t)}\right],
\end{eqnarray}
where the proton wave function $\Psi^{LG}(\mathbf{R},t)$ obeys the free time-dependent Schr\"odinger equation exactly (see the Appendix). Here $\sigma_{\bot}(t)$ and $\sigma_z(t)$ are widths of the wave packet in transverse and longitudinal directions, respectively. \asc{In our theoretical analysis below we will assume that both widths do not depend on time, i.e. $\sigma_{\bot}(t) \equiv \sigma_{\bot} $ and $\sigma_z(t) \equiv \sigma_z$. This is well justified if we compare the characteristic collision time $t_c$ with that of dispersion of the wave packet, $t_d$. Indeed, for a proton wavepacket with a transverse size of sub--$\mu$m, moving with a mean kinetic energy $T_p = 100$~keV, a conservative estimate for the collision time is $t_c \sim 10^{-15}$~s. It is indeed orders of magnitude smaller than the typical dispersion time $t_d \sim 10^{-5}$~s, which can be estimated as:
\begin{eqnarray}
t_d = \sigma_{\perp, z}^2(t = t_0) \, \frac{m_p}{\hbar}, 
\end{eqnarray}
where $m_p$ is the proton mass, see Appendix and Ref.~\cite{Karl2021} for further details.} 

Having briefly discussed the Hamiltonian (\ref{eq:hamiltonian}) of the system, we consider the time--dependent Schr\"odinger equation:
\begin{equation}
    \label{eq:Schroedinger_equation}
    i \hbar \frac{\partial \Psi({\bm r}, t)}{\partial t} = 
    \hat{H} \Psi({\bm r}, t) \, ,
\end{equation}
whose solution $\Psi({\bm r}, t)$ can be constructed in terms of eigenfunctions $\varphi_k(\mathbf{r})$ and eigenvalues $\varepsilon_k$ of the unperturbed Hamiltonian $\hat{H}_0$:
\begin{eqnarray}
   \label{eq:cwavrfunction}
\Psi\,(\mathbf{r},t) = \sum_k a_k(t)\,\varphi_k(\mathbf{r})\exp (-i\varepsilon_k\,t).
\end{eqnarray}
By employing the wave function $\Psi\,(\mathbf{r},t)$ one can obtain the probability 
\begin{eqnarray}
   \label{eq:trans_prob}
\mathcal{P}_f = \left|\left\langle \, \varphi_f \,|\,\Psi\,(\mathbf{r},t)\,\right\rangle\right|^2 = \left|\,a_f\,(t\rightarrow +\infty)\,\right|^2 
\end{eqnarray}
to find the target atom into some specific final state $\ketm{\varphi_f}$ after the collision, i.e. at $t\rightarrow +\infty$.

As seen from Eq.~(\ref{eq:trans_prob}), the analysis of the excitation of the target hydrogen atom by the vortex projectile is traced back to the evaluation of the amplitude $a_f\,(t\rightarrow +\infty)$. In general, this is a very challenging task because of the summation over the entire atomic spectrum in the wave function $\Psi\,(\mathbf{r},t)$. Indeed, $\sum_k$ in Eq.~(\ref{eq:cwavrfunction}) indicates the summation over bound $\ketm{k} = \ketm{n l m}$ and integration over continuum hydrogenic states. The evaluation of the transition amplitude can be greatly simplified, however, within the framework of the first Born approximation:
\begin{eqnarray}
   \label{eq:amplitude}
a_f(t\rightarrow +\infty) \equiv a_{fi} = -i \int_{-\infty}^{\infty}  \exp \left(-i\,\omega_{if}\,t\right) \, \mem{\varphi_f}{\hat{V}(\mathbf{r},t)}{\varphi_i} \,  \mathrm{d}t .
\end{eqnarray}
Here we assume a well defined initial state of the atom, $ \varphi_i = \Psi\,(\mathbf{r},t \to - \infty) $, and introduce the notation $\omega_{if} = \varepsilon_i - \varepsilon_f$ for the transition energy. 

With the help of the first--Born amplitude $a_f(t\rightarrow +\infty)$ we can investigate the excitation $\ketm{i} = \ketm{n_i l_i m_i} \to \ketm{f} = \ketm{n_f l_f m_f}$ of the hydrogen atom due to the Coulomb interaction with the Laguerre--Gaussian ion wave packet. Indeed, by inserting the interaction operator (\ref{eq:intoperator}) into Eq.~(\ref{eq:amplitude}) and by employing the non--relativistic hydrogenic solutions $\varphi_{nlm}({\bm r}) = R_{nl}(r) Y_{lm}(\theta, \phi)$ we found, after some algebra:
\begin{eqnarray}
   \label{eq:amplitude_mod}
    \hspace*{-2cm} a_{fi}({\bm b}) =
    \frac{i \, Z_p \,e^2}{\pi\,v_p\,\sigma^2_{\bot}\,|l_{OAM}|!} \exp \left(-\left(\frac{\omega_{if}\,\sigma_z}{2v_p}\right)^2\right) \sum_{LM} \frac{4 \pi}{2L + 1} \, \mem{l_f m_f}{Y^*_{LM}}{l_i m_i} \,
    I_{LM}({\bm b}) \, ,
\end{eqnarray}
where we introduced the notation:
\begin{eqnarray}
    \label{eq:I_integral}
    I_{LM}({\bm b}) &=& \int {\rm d}\Omega_R \, \Bigg[ \, Y_{LM}(\mathbf{\hat{R}}) \,
    \int {\rm d}R \, {\rm d}r \, R^2 r^2 \,
    \left(\frac{\left(\mathbf{R}_{\bot}-\mathbf{b}\right)^2}{\sigma^2_{\bot}}\right)^{|l_{OAM}|} \nonumber \\[0.2cm]
    && \hspace*{-1cm} \times \frac{r^L_{<}}{r^{L+1}_{>}} R_{n_f l_f}(r) \, R_{n_i l_i}(r)  \exp \left(-\frac{i\,\omega_{if}\, R_z}{v_p}\right)\exp \left(-\frac{\left(\mathbf{R}_{\bot}-\mathbf{b}\right)^2}{\sigma^2_{\bot}}\right)
    \Bigg] \, ,
\end{eqnarray}
with $r_{>} = {\rm max}\left(r, R \right)$ and $r_{<} = {\rm min}\left(r, R \right)$. The evaluation of the four--dimensional integral (\ref{eq:I_integral}) is in general a rather complicated task which can be simplified, however, by using \textit{analytical} solutions $R_{nl}(r)$ of the Scr\"odinger equation for a point--like target nucleus. 
%{\bl Note that we can treat the nucleus as point-like because the transverse coherence length of the incoming proton is supposed to be at least $\sim$ 1 nm, i.e. many orders of magnitude larger that the nucleus size.}

%
%
%
\subsection{Cross section and alignment parameter}
\label{subsec:Cross section}

In previous subsection we have derived the expression for the amplitude  Eq.~(\ref{eq:amplitude_mod}) of the atomic transition, induced by the interaction with a projectile Laguerre--Gaussian wave packet. This amplitude, which depends both on the impact parameter ${\bm b}$ and on the size, velocity and OAM of the wave packet, can be used to derive all observables of the Coulomb excitation process. In the present work, we focus on the $\ketm{1s} \to \ketm{2p}$ excitation whose total probability is given by:
\begin{eqnarray}
   \label{eq:tot_ex_prob}
\mathcal{P}_{\rm tot}({\bm b}) = \sum_{m_f =0,\pm 1} \left|\,a_{2p\,m_f}({\bm b})\,\right|^2 \, ,
\end{eqnarray}
where the summation runs over all magnetic substates $\ketm{2p\,m_f}$ of the final excited state.

Apart from the total excitation probability $\mathcal{P}_{\rm tot}({\bm b})$, we can use the amplitude (\ref{eq:amplitude_mod}) to investigate the magnetic sublevel population of the ion. Most naturally it is described within the framework of the density matrix theory in terms of the so-called alignment parameters $A_{kq}$ \cite{Blu12,BerKa77,BaG00,SurzJPB02}. For an atom in the $p$--state, i.e. with $l_f = 1$, there are in general five alignment parameters $A_{2q}$ with $-2 \leq q \leq 2$ in addition to the trivial one $A_{00}\equiv 1$. While the parameter  
\begin{eqnarray}
   \label{eq:alignment}
    A_{20}({\bm b}) = \sqrt{2} \, \frac{\left|\,a_{2p\,m_f\, =\, \pm1}({\bm b})\,\right|^2-\left|\,a_{2p\,m_f\, = \,0}({\bm b})\,\right|^2}{\left|\,a_{2p\,m_f \,= \,0}({\bm b})\,\right|^2+2\left|\,a_{2p\,m_f \, = \,\pm1}({\bm b})\,\right|^2} \, 
\end{eqnarray}
describes the relative population of the magnetic substates $\ketm{2p \,m_f}$ and is expressed in terms of the transition amplitudes squared, the parameters $A_{2q}$ with $q \ne 0$ characterise the coherence between these substates \cite{SuJ06,Blu12}.

Similar to the total excitation probability $\mathcal{P}_{\rm tot}(b)$, the alignment of the $2p$ state also depends on the position of the target atom $b$ and on the parameters of the incident wave packet. Hence, its measurements can provide more insight into the interaction of atomic targets with vortex ion beams.

\subsection{Angular distribution of fluorescence emission}
\label{subsec:Angular distribution}

After excitation, the excited atomic state will decay into the ground one by the emission of a photon. The angular distribution (as well as the polarization) of this subsequent decay is directly related to the relative population of magnetic sublevels, as described by the set of alignment parameters \cite{Stoeh97,Stoeh99}. For the $\ketm{2p} \to \ketm{1s}$ decay, for example, the angular distribution of emitted radiation can be generally written as:
\begin{eqnarray}
   \label{eq:b_dep_distrib}
    W_b(\theta, \phi)  =  \mathcal{P}_{tot}(b)\left(1 + \sqrt{\frac{4 \pi}{5}} \, 
    \sum\limits_{q} \frac{A_{2q}({\bm b})}{\sqrt{2}} \, Y_{2q}(\theta, \phi) \right)  \, ,
\end{eqnarray}
where the polar $\theta$ and azimuthal $\phi$ angles determine the direction of photon emission with respect to the quantization ($z$--) axis and reaction ($xz$--) plane, respectively. 

The angular distribution (\ref{eq:b_dep_distrib}) is derived for a rather academic case of a single target atom, localized at the distance $b$ from the center of the incident Laguerre--Gaussian wave packet, see Fig.~\ref{Fig1}. In realistic experiments, which can be performed at linear accelerator or ion storage ring, the vortex ions will interact not with a single atom but with a \textit{macroscopic} atomic target. To estimate the angular distribution of the $2p \to 1s$ radiation, emitted from such a target, we have to average the ``individual'' contributions (\ref{eq:b_dep_distrib}) over the impact parameter ${\bm b}$:
\begin{eqnarray}
    \label{eq:floures_emis}
    W_{\rm lab}(\theta) &=& \int W_b(\theta, \phi) \, {\rm d}^2{\bm b} = \tilde{\sigma}_{tot} \left(1 + \frac{\tilde{A}_{20}}{\sqrt{2}} \, P_2(\cos\theta) \right) \, .
\end{eqnarray}
Here we have assumed a homogeneous target whose atoms contribute incoherently to the decay emission. \asc{This assumption is well justified for hot atomic targets that are commonly used at ion accelerators and storage rings \cite{ReB97}. In such targets, collisions between atoms and dispersion of their velocities lead to the loss of coherence of fluorescence radiation from different atoms. In Eq.~(\ref{eq:floures_emis}) we introduced, moreover,} the \textit{effective} excitation cross section $\tilde{\sigma}_{tot}$ and the alignment $\tilde{A}_{20}$. Both $\tilde{\sigma}_{tot}$ and $\tilde{A}_{20}$ are obtained from Eqs.~(\ref{eq:tot_ex_prob}) and (\ref{eq:alignment}), respectively, where the $b$--averaged squared amplitudes $|\tilde{a}_{2p\,m_f}|^2 = \int |a_{2p\,m_f}({\bm b})|^2 \, b \, {\rm d}b$ are employed in place of $|a_{2p\,m_f}({\bm b})|^2$. 

As seen from Eq.~(\ref{eq:floures_emis}), the angular distribution $W_{\rm lab}(\theta)$ of the decay radiation from a macroscopic homogeneous target is defined by the single (effective) alignment parameter $\tilde{A}_{20}$ and does not depend on the azimuthal angle $\phi$. This is an obvious implication of the azimuthal symmetry of the system ``LG ion beam \textit{plus} macroscopic target'', and can be trivially obtained by integrating Eq.~(\ref{eq:b_dep_distrib}) over $\phi_b$ when using the standard transformation of the alignment parameters $A_{2q}({\bm b}) = {\rm e}^{i q \phi_b} A_{2q}(b)$, with $q = 0, \pm 1, \pm 2$, see Ref.~\cite{Mai2020} for further details. 

%----------------------------------------- Results ----------------------------------------------%
\section{Results and discussion}
\label{sec:results}

In the present work, we focus on the $1s \to 2p$ excitation and subsequent radiative decay of the hydrogen target in collisions with the Laguerre--Gaussian proton beam, i.e. for $Z_p = 1$. Our special interest to the $\mathrm{H}(1s) + \mathrm{H}^{+}$ collisions stems from the fact that they have been extensively studied in the past for the conventional \textit{plane--wave} projectiles. \asc{Based on the well--established plane--wave results, we first discuss below the validity of our theoretical model.} 

%The well--established  plane--wave results will serve as a benchmark for our calculations and will help us to better emphasize the effect of ion's OAM in Coulomb excitation of the target atoms. 

\asc{As mentioned already in Section~\ref{subsec:Excitation amplitude}, our analysis of the Coulomb excitation will be performed within the framework of the first--Born approximation. Using this rather simple approach will allow us to understand and illustrate the main qualitative effects induced by OAM of vortex ions. Moreover, the application of the Born approximation is well justified for proton kinetic energies of 100 keV and above, which are likely to be exploited in future experiments and, hence, which are of main interest for the present study. For such energies, corresponding to mean projectile velocities $v_p \gtrsim 2$~a.u., the first Born calculations are known to reproduce well the predictions based on more sophisticated models for most impact parameters, except for very small $b$, see Refs.~\cite{Bates61, SaT96} for further details. }

\asc{Discussing the details of the theoretical model used, one can also note that the $1s \to 2p$ excitation of a hydrogen atom is a special case due to the (near) degeneracy of $2s$ and $2p$ levels. This degeneracy might lead to a noticeable coupling to the $2s$ state and hence to the virtual $1s \to 2s \to 2p$ transition, affecting the excitation probabilities $a_{2p m_f}$. The channel coupling is most pronounced, however, near excitation threshold and can be neglected for relatively high collision energies, for which the first Born approximation is valid \cite{Bates61, SaT96, Tan03}. Since the present study is focused on the high--energy (Born) regime, the $2s$ coupling will not be considered below.}

\begin{figure}[t]
	\includegraphics[width=0.8\linewidth]{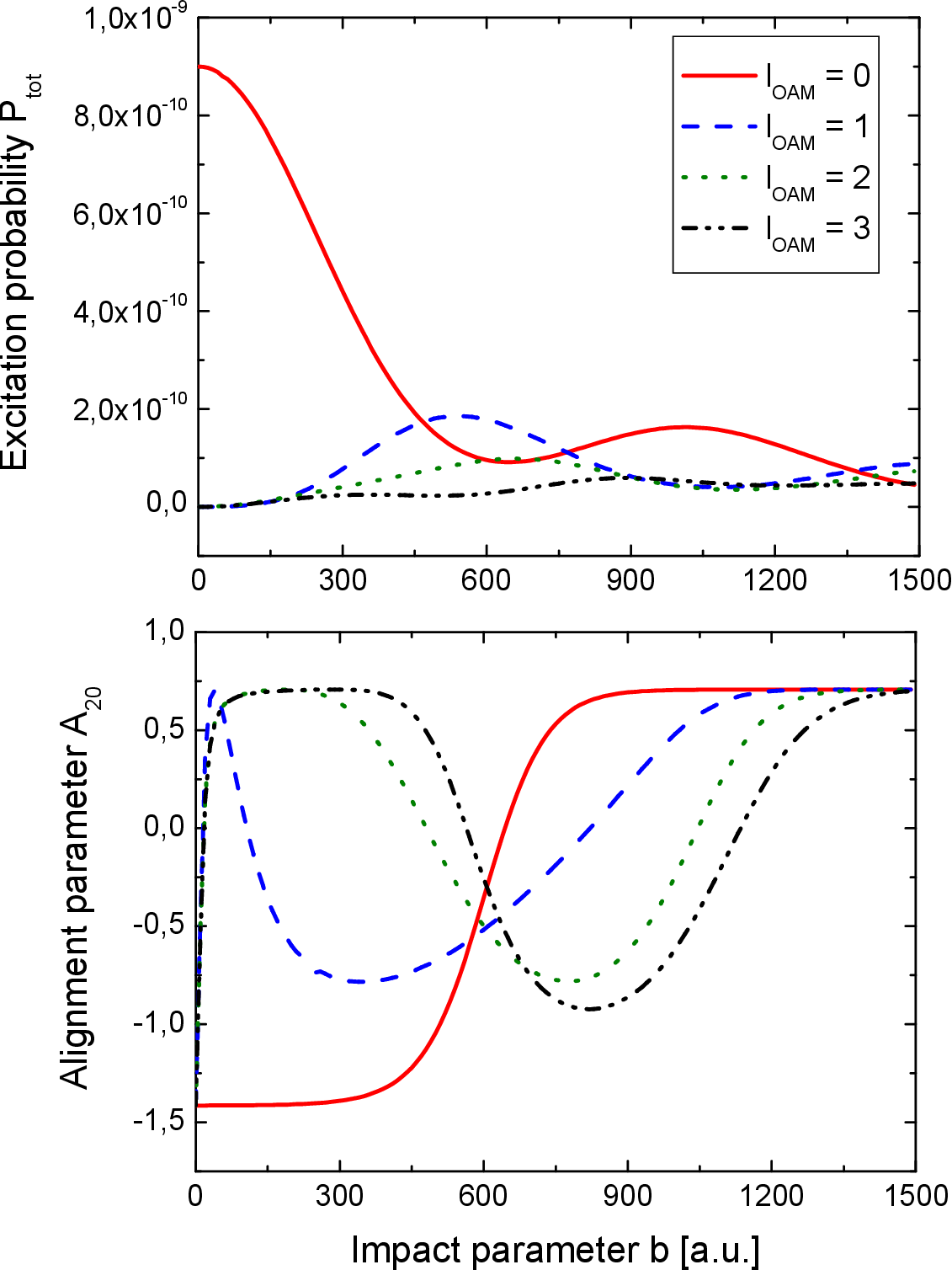}
%    \vspace*{-1cm}
	\caption{Total excitation probability (top) and alignment parameter (bottom) for Coulomb excitation of the hydrogen atom by the vortex proton wave packet with the \asc{mean kinetic energy $T_p = 100$~keV, the width $\sigma_\perp = 500$ a.u.} and various OAM projections: $l_{OAM} = 0,1,2,3$, as functions of the impact parameter $b$.}
    \label{Fig2}
\end{figure}

\asc{Having briefly justified our theoretical approach, we turn to the main question of the present study, the sensitivity of the $1s \to 2p$ Coulomb excitation to the OAM projection of the incident wave packet.} We start with the case of a single hydrogen atom, being initially in the ground $1s$ state and displaced by the distance $b$ from the axis of the Laguerre Gaussian beam (\ref{eq:chargedensity}). For this case, we display in Fig.~\ref{Fig2} the total excitation probability $\mathcal{P}_{\rm tot}(b)$ (top panel) and the alignment parameter $A_{20}(b)$ of the $2p$ state (bottom panel). The results are presented for the incident wave packet with the \asc{mean kinetic energy $T_p = 100$ keV, the transverse width $\sigma_{\perp} = 500$ a.u., and with four different OAM projections: $l_{OAM} = 0$ (red solid line), $l_{OAM} = 1$  (blue dashed line), $l_{OAM} = 2$ (green dotted line), and $l_{OAM} = 3$ (black dash--dotted line)}. As follows from the figure, both the excitation probability and the alignment parameter strongly depend on the OAM projection of the vortex ion beam. In particular, one can observe a substantial decrease of $\mathcal{P}_{\rm tot}(b)$ for higher OAM projections \asc{at relatively small impact parameters, $b \lesssim 300$~a.u.. It can be explained} by qualitatively distinct charge densities (\ref{eq:chargedensity}) of the wave packets with $l_{OAM} = 0$ and $l_{OAM} \ne 0$: while the $\rho(l_{OAM} = 0)$ is maximal at $b = 0$, the $\rho(l_{OAM} \ne 0)$ exhibits the ring structure with the low--density spot around the beam center. \asc{The annular density structure of projectiles, carrying non--zero OAM, is also reflected in the position of maxima of $\mathcal{P}_{\rm tot}(b)$ for various $l_{OAM} \ne 0$. Indeed, Eq.~(\ref{eq:chargedensity}) implies that the size of the charge--density ring of LG wave packet increases with the growth of $l_{OAM}$, thus leading to a shift of the maximum of the excitation probability towards larger $b$.}

\begin{figure}[t]
	\includegraphics[width=1.0\linewidth]{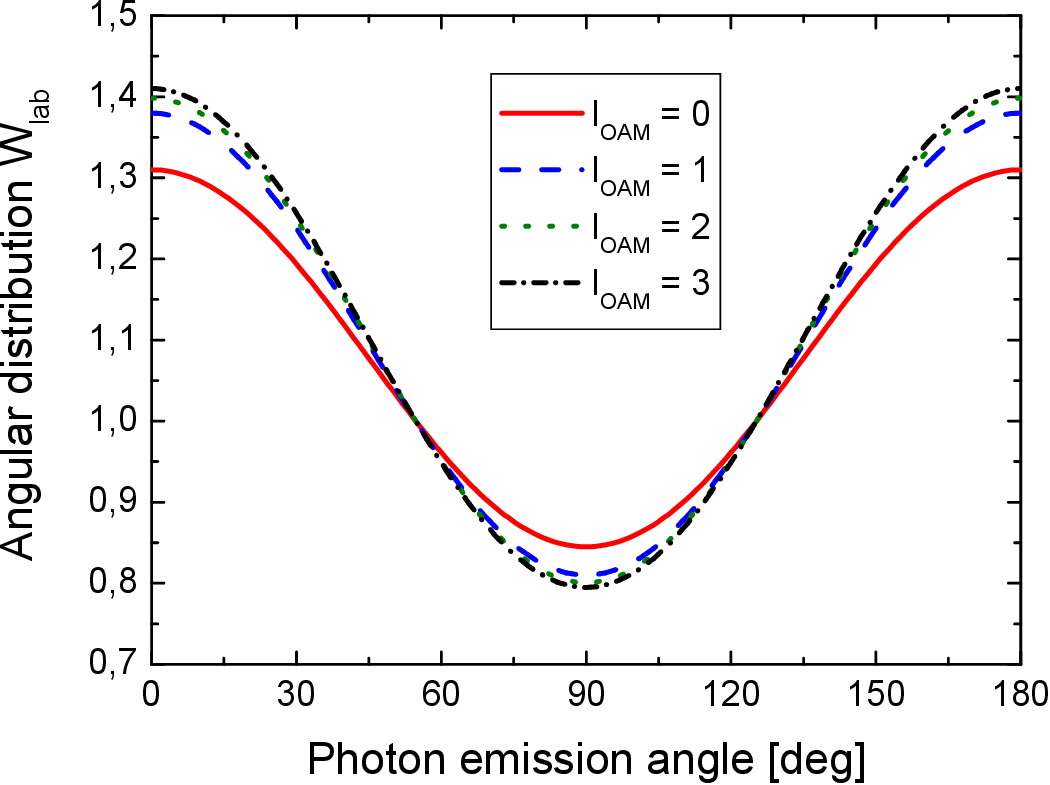}
	\vspace*{-0.5cm}
	\caption{Angular distribution of the $2p \to 1s$ radiative decay following Coulomb excitation of hydrogen atoms by Laguerre--Gaussian proton beam. Since we are primarily interested in the \textit{shape} of the angular distribution, $W_{lab}(\theta)$ is normalized here as $\int W_{lab}(\theta) \, {d}\Omega = 1$. The characteristics of the ion beam are the same as for Fig. \ref{Fig2}. 
	}
    \label{Fig3}
\end{figure}

Besides the total excitation probability, the alignment parameter $A_{20}(b)$ also exhibits a remarkable sensitivity to the OAM projection of projectile ions. \asc{Most clearly this sensitivity is manifested for impact parameters $b \lesssim$~1200~a.u., for which the presence of a non--zero OAM projection of the projectile wave packet can even lead to a \textit{sign reversal} of the alignment parameter $A_{20}(b)$. Even though $A_{20}$'s for various $l_{OAM}$ approach each other for $b > 1500$~a.u., such an OAM--induced ``alignment inversion'' at (relatively) small $b$ can significantly affect the angular properties of subsequent fluorescence radiation for a realistic scenario of macroscopic target. As will be shown below, this will open up the opportunity to diagnose OAM of vortex ion beams.}

% ------------------------------------ Figure 4 -------------------------------------------
%
\begin{figure}[t]
	\includegraphics[width=0.9\linewidth]{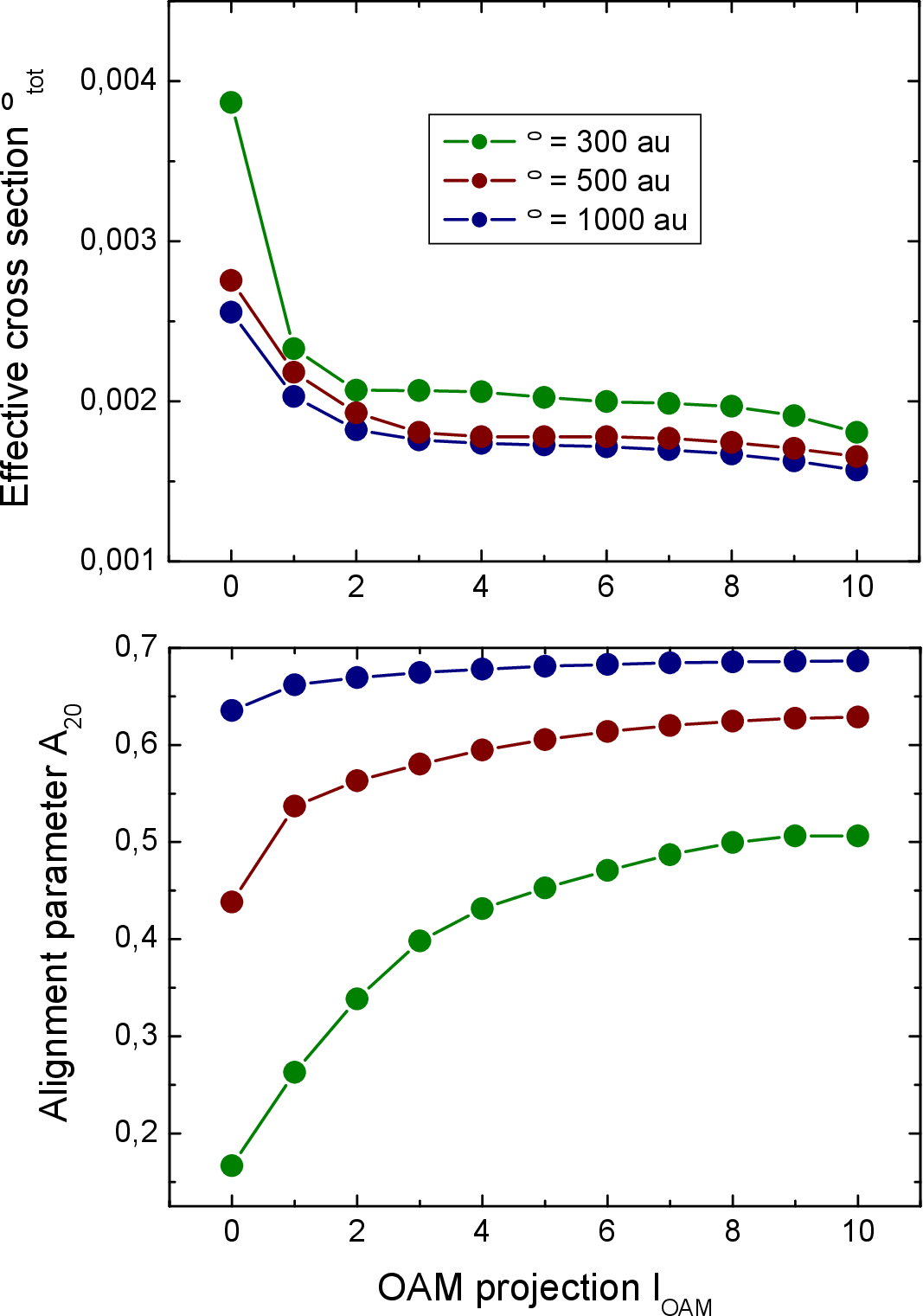}
%	\vspace*{-1cm}
	\caption{Effective excitation cross section $\tilde{\sigma}_{tot}$ (top panel) and alignment parameter $\tilde{A}_{20}$ (bottom panel) for the macroscopic hydrogen target, bombarded by the proton wave packet with the \asc{kinetic energy $T_p = 100$~keV and transverse widths $\sigma_{\perp}$~=~300 a.u. (green circles), 500~a.u. (maroon circles) and 1000~a.u. (blue circles)}. Results are presented for various OAM projections.
	}
    \label{Fig4}
\end{figure}

\asc{After discussing the excitation of individual atoms, located at a certain distance $b$ from the center of a projectile wave packet, we are ready to consider the experimentally relevant scenario of \textit{macroscopic} atomic target, bombarded by LG protons. For this case we will pay a special attention to the angular distribution of subsequent $2p \to 1s$ radiative decay. The angle--resolved measurements of photoemission from (collisionally--excited) target atoms is one of the best--established methods to investigate details of a collision process \cite{BaG00,BerKa77,Blu12}.} While in the past, this method has been widely applied for conventional plane--wave projectiles, here we propose to extend it for the diagnostics of Laguerre--Gaussian beams. In order to advocate the feasibility of such (future) experiments, we display in Fig.~\ref{Fig3} the angular distribution of $2p \to 1s$ radiation, emitted from the \textit{macroscopic} hydrogen target bombarded by protons with $l_{OAM} = 0$ (red solid line), $l_{OAM} = 1$ (blue dashed line), $l_{OAM} = 2$ (green dotted line) and $l_{OAM} = 3$ (black dash-dotted line). The angular distribution was calculated based on Eq.~(\ref{eq:floures_emis}), \asc{for the transverse beam's spread $\sigma_{\perp} =$~500~a.u. and mean kinetic energy $T_p = 100$~keV}. As seen from the figure, the anisotropy of the $2p \to 1s$ emission is rather sensitive to the $l_{OAM}$ of projectile ions. \asc{Namely, it is lowest for a wave packet carrying no orbital angular momentum, $l_{OAM} = 0$, and becomes more pronounced with increasing $l_{OAM}$. This variation of the (anisotropy of) angular distribution for different OAM projections can easily observed with the modern detectors and can open a route for the diagnostics of vortex ion beams.}

\asc{As follows from Eq.~(\ref{eq:floures_emis}), the sensitivity of the angular anisotropy of $2p \to 1s$ radiation to the OAM of the projectile wave packet reflects that of the effective alignment parameter $\tilde{A}_{20}$. To better understand the latter, Fig.~\ref{Fig4} shows $\tilde{A}_{20}$ for the OAM projections in the range from $l_{OAM}$~=~0, which corresponds to the non--vortex Gaussian beam, to $l_{OAM}$~=~10. Moreover, calculations have been performed for various lateral widths of the wave packet: 300 a.u. (green circles), 500~a.u. (maroon circles) and 1000~a.u. (blue circles). For all these packet sizes, the effective alignment parameter grows as the $l_{OAM}$ increases, thus leading to a stronger anisotropy of the $2p - 1s$ emission as noted above. One can note, that the OAM--sensitivity becomes less pronounced at larger transverse widths $\sigma_{\perp}$ of the wave packet. However, even for the largest packet of $\sigma_\perp =$~1000~a.u., considered here, the variation of the alignment parameter between $\tilde{A}_{20}(l_{OAM} = 0)$~=~0.63 to $\tilde{A}_{20}(l_{OAM} = 10)$~=~0.69 can be detected in modern experiments and can be used for the diagnostics of LG ion beams.}

%
%-------------------------------------------------- Summary and outlook ------------------------------------------------------------%
\section{Summary}
\label{sec:outlook}
In this work we have theoretically studied the Coulomb excitation of one--electron atoms by Laguerre--Gaussian ionic wave packets. Special attention was paid to the question of how the population of the magnetic sublevels of the excited atom is affected by the OAM projection of the projectile ions. In order to explore this OAM--dependence, we have applied the non--relativistic first--Born approximation and the density matrix theory. The general expressions were derived for the total excitation probability and the alignment parameter of the excited atom. Based on the theory obtained, detailed calculations have been carried out for the $\mathrm{H}(1s) + \mathrm{H}^{+}_{LG}\rightarrow \mathrm{H}(2p) + \mathrm{H}^{+}_{LG}$ excitation of hydrogen atoms by vortex proton beam. In these calculations, we have addressed not only a rather academic case of the excitation of a single well--localized atom, but also the much more promising---from the experimental viewpoint---scenario of collision of the ion beam with the macroscopic hydrogen target. For the latter case, particular emphasis was placed on the angular distribution of the subsequent $2p \to 1s$ radiative emission from the whole target, which is defined by the magnetic sublevel population of excited atoms and which can be easily measured in modern experiments. We have shown that the anisotropy of the $2p \to 1s$ radiation and, hence, the alignment of the $2p$ state are rather sensitive to the OAM projection of the Laguerre--Gaussian beam. This sensitivity may open up a promising avenue for the diagnostics of vortex ions in accelerator and storage ring experiments. \asc{To fully explore the potential of Coulomb excitation for the ion diagnostics, however, additional analysis is needed for even wider wave packets, $\sigma_{\perp} >$~1~$\mu$m, and to elucidate the role of their longitudinal size $\sigma_{z}$. Such an analysis, being computationally demanding and going beyond the scope of the present proof--of--principle study, will be presented in the forthcoming publications.}

%-------------------------------------------------- Acknowledgments  ------------------------------------------------------------%
\ack
\label{sec:acknowledgements}
The publication is funded by the Open Access Publishing Fund of GSI Helmholtzzentrum
f\"{u}r Schwerionenforschung.
A.S. acknowledges support by the Deutsche Forschungsgemeinschaft (DFG, German Research Foundation) under Germany’s Excellence Strategy – EXC-2123 QuantumFrontiers – 390837967.

\section*{Appendix: wave function of the twisted proton}

%The wave function of the twisted proton obeys the free Schr\"odinger equation. Along with the well-known plane wave, its exact solution is the so-called standard \textit{Laguerre-Gaussian wave packet} \cite{Karl2021} (in the following we will assume $\hbar = c = 1$)

Apart from the well--known plane wave, the Laguerre-Gaussian wave packet:
%
%
%\begin{widetext}
\begin{eqnarray}
    \label{LGpsi}
 \hspace{-1.5cm} \Psi^{LG}(\bR, t) = \sqrt{\frac{n!}{(n + |\ell_{OAM}|)!}}\, \frac{1}{\pi^{3/4}}\left(\frac{R_{\perp}}{\sigma_{\perp}(t)}\right)^{|\ell_{OAM}|}\frac{1}{\sigma_{\perp}(t)\sqrt{\sigma_z(t)}}\, L_{n}^{|\ell_{OAM}|}\left(\frac{R_{\perp}^2}{\sigma_{\perp}^2(t)}\right) \nonumber \\  \hspace{-1.5cm}
 \times \exp\Big\{-\frac{i}{\hbar}t\frac{\langle p\rangle^2}{2m_p} + \frac{i}{\hbar}\langle p\rangle R_z + \frac{i}{\hbar}\ell_{OAM}\phi_r - i (2n + |\ell_{OAM}| + 1)\arctan (t/t_{d,\perp})  \nonumber \\ \hspace{-1.5cm} - \frac{i}{2}\arctan (t/t_{d,z})
  - \frac{R_{\perp}^2}{2\sigma_{\perp}^2(t)}\, (1-it/t_{d,\perp}) - \frac{(R_z-v_p t)^2}{2\sigma_z^2(t)}\, (1-it/t_{d,z})\Big\} \, ,
\end{eqnarray}
%\end{widetext}
% убрал из числителя i^{2n+\ell_{OAM}}
%
with the standard normalization
\begin{equation}
    \int d^3 R\, |\Psi^{LG}(\bR,t)|^2 = 1 \, ,
\end{equation}
is also an exact solution of the free--particle Schr\"odinger equation. Here, $\bR = \{R_{\perp} \cos{\phi_r},R_{\perp} \sin{\phi_r}, R_z\}$, $m_p$ is the proton mass, $\langle p\rangle = v_p\,m_p$ is a mean momentum, $n=0,1,2,..$ is a principal quantum number and $l_{OAM} = 0, \pm 1, \pm 2, ...$ is a projection of the orbital angular momentum onto the propagation axis. 

The set of solutions (\ref{LGpsi}) for $n = 0, 1, 2 ...$ and $l_{OAM} = 0, 1, 2...$ is \textit{complete and orthogonal}. It describes, moreover, free rectilinear motion on average, 
\bea
& \displaystyle \langle R_z\rangle(t) = v_p\, t,
\eea
and spreading of the wave packet with time both in the transverse and longitudinal directions. For example, for the transverse direction one can find: 
\begin{eqnarray}
\sigma^2_{\perp}(t) &=& \frac{\hbar^2}{(\sigma^{p}_{\perp})^2} \left(1 +\frac{t^2}{t_{d,\perp}^2}\right) = \sigma^2_{\perp}(\,t_0)+ \left(\frac{\sigma^p_{\perp}}{m_p}\right)^2 t^2, \nonumber \\ 
\sigma_{\perp}(t_0) &=& \frac{\hbar}{\sigma^p_{\perp}},\  
\end{eqnarray}
where 
\asc{
\begin{equation}
t_{d,{\perp}} = \frac{m_p\hbar}{(\sigma^p_{\perp})^2} = \frac{m_p \sigma^2_{\perp}(t_0)}{\hbar}
\end{equation}
is the time after which $\sigma_{\perp}(t_d) = \sqrt{2} \sigma_{\perp}(t_0)$, and $\sigma^p_{\perp}$} is the width of the wave packet in the momentum space. Similar expressions can be obtained for the longitudinal direction:
\bea
& \displaystyle \sigma^2_z(t) = \frac{\hbar^2}{(\sigma^p_z)^2} \left(1 + \frac{t^2}{t_{d,z}^2}\right) = \sigma^2_z(t_0) + \left(\frac{\sigma^p_z}{m_p}\right)^2 t^2,\cr 
& \displaystyle \sigma_z(t_0) = \frac{\hbar}{\sigma^p_z},\ t_{d,z} = 
\frac{m_p\hbar}{(\sigma^p_z)^2} = \asc{\frac{m_p \sigma^2_z(t_0)}{\hbar}}.
\eea
One can note that for both cases $t_{d,{\perp}}$ and $t_{d,z}$ are inversely proportional to \asc{$\hbar/m_p \approx 6 \times 10^{-8}$~m$^2$/s, which implies characteristic dispersion time of about $t_{d} \sim 10^{-5}$~s} for sub--$\mu$m--size proton wave packets.

%Let us suppose for simplicity that $\sigma_z=\sigma_{\perp}=\sigma$. %Then we find the following root-mean-square values:
%\bea 
%& \displaystyle \sqrt{\langle {\bm u}_{\perp}^2\rangle} = %\frac{\sigma}{m}\, \sqrt{2n + |\ell| + 1},\cr
%& \displaystyle \sqrt{\langle \rho^2\rangle(0)} = \frac{1}{\sigma}\, %\sqrt{2n + |\ell| + 1}.
%\eea
%This wave packet spreads with time according to the general law
%\bea
%& \displaystyle \langle\rho^2\rangle(t) = \langle\rho^2\rangle(0) + %\frac{\partial \langle\rho^2\rangle(0)}{\partial t}\, t + \langle{\bm %u}_{\perp}^2\rangle\, t^2,
%\eea
 %If the focus of the packet is in the initial moment of time, i.e. %$\partial \langle\rho^2\rangle(0)/\partial t=0$, then  
%\bea
%& \displaystyle \langle\rho^2\rangle(t) = \langle\rho^2\rangle(0) + %\langle{\bm u}_{\perp}^2\rangle\, t^2 = \langle\rho^2\rangle(0) %\left(1 + \frac{t^2}{t_d^2}\right) = \langle\rho^2\rangle(0) \left(1 %+ \frac{\langle z\rangle^2}{z_R^2}\right),
%\eea
%where $t_d = \sqrt{\langle\rho^2\rangle(0)/\langle{\bm %u}_{\perp}^2\rangle}$ is the diffraction (spreading) time and $z_R = %\langle u\rangle t_d$ is the Reyleigh length. When spreading is %essential, $\langle z\rangle \gg z_R$, we have 
%\bea
%& \displaystyle \langle\rho^2\rangle(\langle z\rangle) \approx %\langle\rho^2\rangle(0) \frac{\langle z\rangle^2}{z_R^2},\cr 
%& \displaystyle \langle z\rangle = \langle u\rangle %\sqrt{\frac{\langle\rho^2\rangle(\langle z\rangle)}{\langle{\bm %u}_{\perp}^2\rangle}} = \frac{\langle p\rangle}{2n + |\ell| + 1} %\sqrt{\langle \rho^2\rangle(\langle z\rangle)} %\sqrt{\langle\rho^2\rangle(0)}
%\eea
At large distances from the source compared to the Rayleigh length, the root-mean-square radius of the wave packet, which can be written in terms of the transverse spreading of the wave packet as $\langle R_{\perp}^2\rangle(t) = \sigma^2_\perp(t)\,(2n + |l_{OAM}| + 1)$, spreads as \cite{Karl2021}:
\bea
& \displaystyle \sqrt{\langle R_{\perp}^2\rangle(t)} \equiv \langle R_{\perp}\rangle = \lambda\, \frac{\langle R_z\rangle}{ \langle R_{\perp}\rangle_0}\, (2n + |l_{OAM}| + 1),
\eea
with $\lambda = \hbar/\langle p\rangle$ is the de Broglie wavelength and $\langle R_{\perp}\rangle_0\equiv \sqrt{\langle R_{\perp}^2\rangle(t\,=\,t_0)}$. For the Gaussian mode with $n = 0$ and $l_{OAM}=0$, this is simply the far-field \textit{van Cittert–Zernike theorem}.

%
%
%--------------------------------------Bibliography-----------------------------------------------------------------
%
%
\section*{References}
\bibliographystyle{iopart-num.bst}
\bibliography{main.bib}
\end{document}